\documentclass[reprint, amsmath,amssymb, aps,superscriptaddress,nofootinbib,showkeys]{revtex4-2}
\usepackage{graphicx}
\usepackage{amsmath,amssymb,amsfonts}
\usepackage{amsthm}
\usepackage{xcolor}
\usepackage{textcomp}
\usepackage{booktabs}
\usepackage{algorithm}
\usepackage{algorithmicx}
\usepackage{algpseudocode}
\usepackage{listings}
\usepackage[utf8]{inputenc}
\begin{document}

\title{Shortcut to adiabatic transfer in a four-level N-type system using counterdiabatic driving}


\author{Marius Romuald Kamsap}\email{mkamsap@yahoo.com}\author{Simon Evehe Ndong}\author{Ma\"ik Delon Mboumba}
\affiliation{Laboratoire d'Optique, Laser et Applications, Université des Sciences et Techniques de Masuku, Franceville, Gabon}
\affiliation{Département de Physique, Faculté des Sciences, Université des Sciences et Techniques de Masuku, Franceville, Gabon}
\author{Mathurin Esouague  Ateuafack}%\email{iiauthor@gmail.com}
\affiliation{Department of Electrical and Electronics Engineering, College of Technology, University  of  Buea, Buea, Cameroon}


\author{Christian Kenfack-Sadem}%\email{iiiauthor@gmail.com}
\affiliation{Unit\'e de Recherche de Mati\`ere Condens\'ee, d'Electronique et de Traitement du Signal, Department of Physics, Faculty of Science,  University of Dschang, Dschang, Cameroon}

\author{Thierry Blanchard  Ekogo}%\email{iiiauthor@gmail.com}
\affiliation{Laboratoire d'Optique, Laser et Applications, Université des Sciences et Techniques de Masuku, Franceville, Gabon}
\affiliation{Département de Physique, Faculté des Sciences, Université des Sciences et Techniques de Masuku, Franceville, Gabon}
\author{Alain-Brice Moubissi}%\email{iiiauthor@gmail.com}
\affiliation{Laboratoire d'Optique, Laser et Applications, Université des Sciences et Techniques de Masuku, Franceville, Gabon}
\affiliation{Département de Physique, Faculté des Sciences, Université des Sciences et Techniques de Masuku, Franceville, Gabon}
\author{Caroline Champenois}

%\affiliation[2]{ D\'epartement de Physique, Facult\'e des Sciences, Universit\'e des Sciences et Techniques de Masuku, Franceville, Gabon}

\affiliation{Aix-Marseille Universit\'e,CNRS, PIIM, Marseille, France}


\begin{abstract}
The stimulated Raman shortcut-to-adiabatic passage (STIRSAP) technique has been demonstrated to accelerate adiabatic population transfer processes. Traditionally applied to three-level atomic systems, this method enables rapid and efficient population transfer in a short interaction time. In this work, we extend this approach to a four-level $N$-type system driven by three laser fields. Under the three-photon phase matching condition, the first-order Doppler effect can be completely eliminated, making this configuration more suitable for applications in frequency metrology and quantum information processing. By adiabatically eliminating the intermediate state, the four-level system is effectively reduced to a two-level model, enabling the application of counterdiabatic driving without requiring additional couplings. This approach allows for fast and efficient population transfer between metastable states. We analyze the influence of laser intensity peaks and detuning, and we show that the transfer time is significantly shorter than that achieved with the STIRAP technique. Moreover, we also show that an optimal coupling strength between the ground and metastable states further minimizes the operation time.
\end{abstract}



\maketitle
\section{Introduction\label{intro}}
%\cancel{ddddd}
In recent decades, quantum manipulation and coherent population transfer between internal atomic states have attracted significant attention in both theoretical and experimental research \cite{shore91, yatsenko97, li09, giannelli14, li16, vitanov17, li17, blekos20, li25}. The coherent superposition of quantum states plays a crucial role in a wide range of scientific disciplines, including quantum optics, quantum chemistry \cite{sterling96, janszky93}, mechanical manipulation, and laser cooling of atoms \cite{bakos96}. To achieve efficient and reliable coherent atomic population transfer, several techniques have been developed. These include resonant pulse methods \cite{allen87} and adiabatic approaches such as the chirped laser method \cite{band94, ignacio99, djotyan00, kamsap25}, rapid adiabatic passage (RAP), and stimulated Raman adiabatic passage (STIRAP) \cite{Bergmann98, Kuklinski89}. The resonant pulse method is known for its speed and effectiveness; however, it requires high-intensity laser pulses and exhibits strong sensitivity to parameter fluctuations. By contrast, adiabatic approaches are more robust and stable because they rely on slowly varying system parameters to maintain adiabatic following. However, this advantage comes at the cost of longer interaction times, which limits their practical speed.  

In quantum information processing, it is crucial for the process to be extremely fast in order to minimize the influence of the environment, which leads to decoherence in the system \cite{feng08, singhal25}. This necessitates the development of robust and fast control techniques that operate with moderate laser intensities. To address this requirement, methods based on optimal control theory \cite{sola99, sugny08, vasilev09} and composite pulses \cite{torosov11, torosov13} have been proposed. These approaches reduce operation times, minimize systematic errors, and mitigate unwanted losses. More recently, extensive efforts have been devoted to shortcuts to adiabaticity (STA), including counterdiabatic (CD) which replicate the dynamics of adiabatic population transfer within a much shorter time scale \cite{drimpak03, chen10, chen11, chen12, deam13, martinez14, baksic16, li16}. One notable example is the stimulated Raman shortcut-to-adiabatic passage (STIRSAP) \cite{li16}, which relies on counterdiabatic drivingemploys counterdiabatic driving to suppress diabatic transitions.

In counterdiabatic driving \cite{drimpak03, berry09,  chen10, campo13}, an auxiliary control field is introduced such that the system exactly follows the adiabatic reference trajectory, thereby eliminating nonadiabatic transitions. This method has been demonstrated both theoretically and experimentally in two-level systems, and later extended to three-level atomic systems in a $\Lambda$ configuration \cite{chen10, bason12, zhang13}. In the three-level case, however, implementing an additional coupling between the initial and target states often realized via magnetic dipole transitions leads to a triangular configuration, which can be experimentally challenging or even unfeasible in certain setups. A common workaround involves applying a unitary transformation to remove the unwanted coupling while preserving equivalent system dynamics \cite{song16}. Nonetheless, a persistent issue remains: the Doppler effect. Three-level systems interacting with two laser fields are inherently sensitive to thermal atomic motion, as the Doppler shift disturbs the precise two-photon resonance condition required for STIRAP. Maintaining perfect two-photon resonance becomes even more difficult in multiphoton transitions. To overcome this, three-photon resonance schemes using four-level atomic systems in an $N$-type configuration have been proposed \cite{Champenois06, barker16}, effectively mitigating the Doppler-induced frequency shifts.

The four-level atomic system in the $N$-configuration considered in this work corresponds to the lowest energy levels of the $^{40}$Ca$^+$ ion (see Fig.~\ref{fig_N}). A similar level configuration can also be realized in alkaline-earth atoms with hyperfine structure or in other alkali-like ions. The system consists of a three-level $\Lambda$ subsystem formed by two electric dipole transitions, to which a weak quadrupole coupling is added between the ground and metastable level, resulting in the complete $N$-type configuration.

In this system, three-photon coherent population trapping (CPT) has been identified and analyzed in Ref.~\cite{Champenois06}. This phenomenon happens on a Raman like three photon resonance condition and produces a narrow dark line in the fluorescence spectrum, which can be exploited as a terahertz frequency standard in an ion cloud \cite{champenois07}. When the three-photon  phase matching condition is satisfied, the first-order Doppler effect can be exactly canceled from the CPT resonance condition due to the geometric configuration of the laser fields. We have previously demonstrated the efficiency of population transfer between the two metastable states using the STIRAP technique in this system \cite{kamsap13}.

In this paper, we focus on the implementation of stimulated Raman shortcut-to-adiabatic passage in four-level quantum systems. We develop an analytical approach showing that, in such systems, the lasers satisfying the dark-resonance condition enable the cancellation of the first-order Doppler effect. The four-level system in the $N$-configuration can be reduced to an equivalent three-level $\Lambda$ configuration, where STIRAP applies naturally. By performing adiabatic elimination of the intermediate states, we further obtain an effective two-level system. This reduction allows us to apply counterdiabatic driving together with a unitary transformation previously proposed for two-level systems, thereby realizing STIRSAP without introducing any additional coupling fields simply by modifying the temporal shapes of the laser pulses. Moreover, we show that the peak laser intensity, pulse rate, and detuning play critical roles in determining both the transfer efficiency between the metastable states and the operation time of the protocol.

This manuscript is organized as follows. Section II presents the implementation of STIRAP in the four-level system, introducing the concept of coherent population trapping under three-photon resonance. Section III describes the STIRSAP protocol, based on counterdiabatic driving. Section IV reports the numerical simulations modeling population transfer between stable and metastable states and discusses optimization strategies to minimize operation time. Finally, Section V provides concluding remarks.

 \section{Implementation of STIRAP in the four levels $N$-system}

\begin{figure}[htbp]
\includegraphics[scale=0.45]{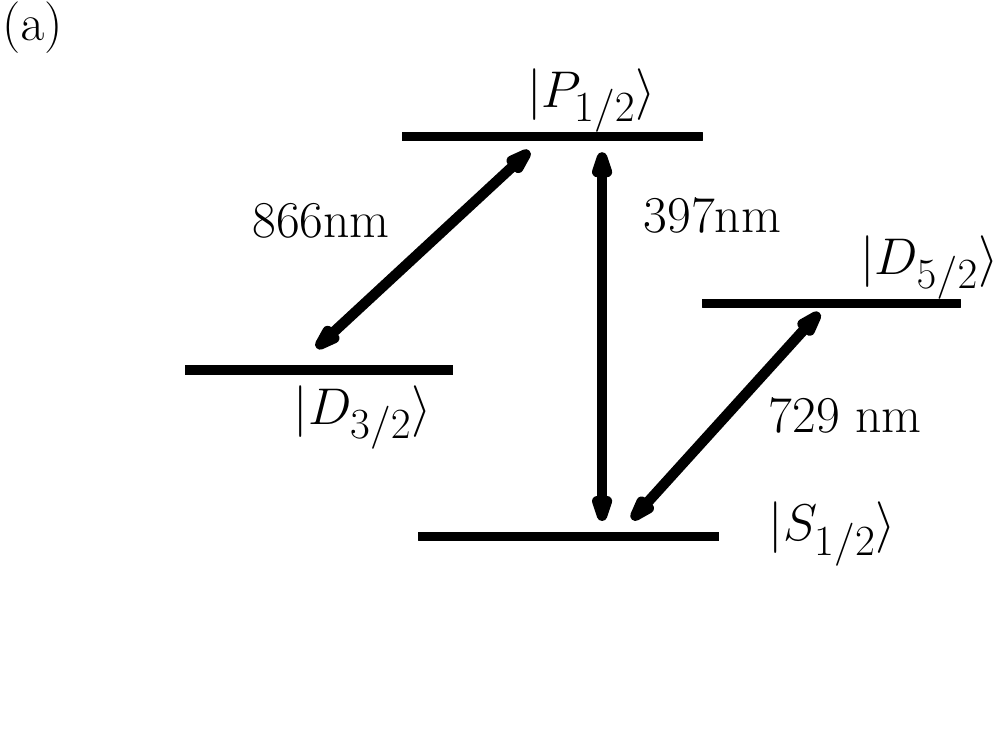}
\includegraphics[scale=0.45]{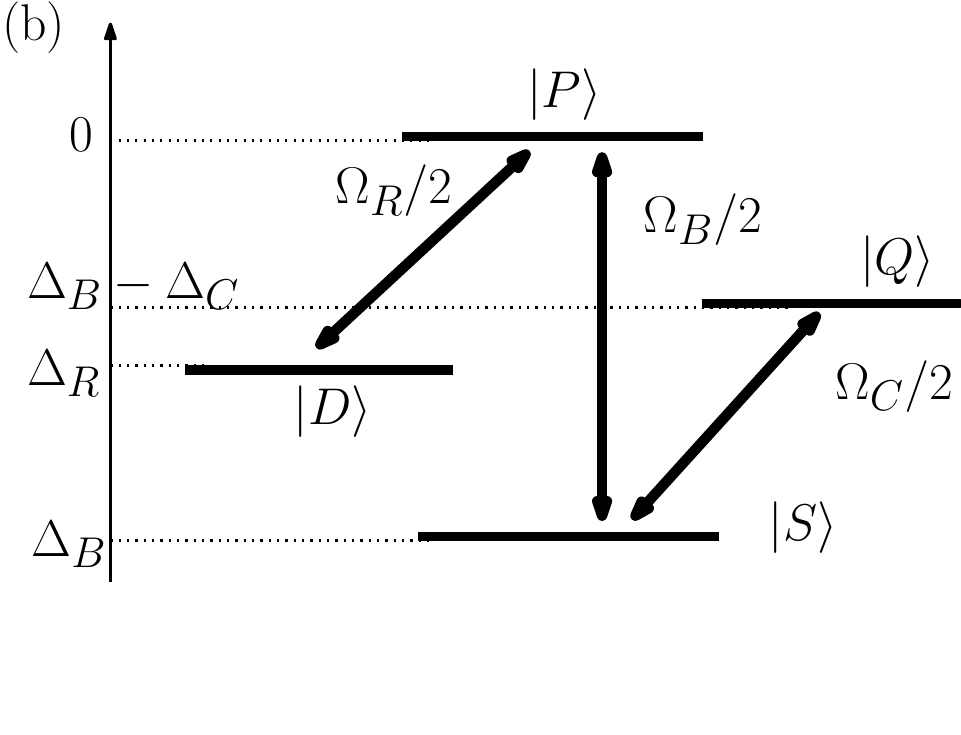}
\caption{ (a): Transition scheme for the three-photon CPT in Ca$^+$, (b):  the equivalent energy diagram of $N$ scheme in the dressed state picture. \label{fig_N}}
\end{figure}

The atomic system under consideration is the $^{40}$Ca$^+$. However, an identical configuration of energy levels can also be realized in alkaline-earth atoms possessing hyperfine structure, as well as in other alkali-like ions with a metastable $d$-orbital, such as Hg$^+$, Ba$^+$, Sr$^+$. The $N$-shaped scheme, illustrated in  Fig.~\ref{fig_N}-\textbf{a} involves three stable or metastable states:
$|S_{1/2}\rangle$,  $|D_{3/2}\rangle$, and $|D_{5/2}\rangle$, which we denote respectively as $|S\rangle$, $|D\rangle$ and $|Q\rangle$. The $|S_{1/2}\rangle$ and  $|D_{3/2}\rangle$ are each coupled to the short-lived excited state $|P_{1/2}\rangle$ state (denoted $|P\rangle$) by electric dipole transitions, labeled B and R in the figure, with respective wavelengths of 397nm and 866nm.  In addition, the transition $|S\rangle\to|Q\rangle$, marked by "C" in Fig.~\ref{fig_N}-\textbf{b}, is an electric quadrupole transition with a wavelength of 729 nm. The associated coupling strength is typically much weaker than that of the electric dipole transitions, reflecting the higher-order nature of the quadrupole interaction.
 We denote by $\omega_{ij}$ ($i,j=S,P,D,Q$) the resonance frequencies of the transitions. Transitions $|S\rangle\to |P\rangle$, 
$|D\rangle\to |P\rangle$ and $|S\rangle\to|Q\rangle$ are coupled by lasers at frequency $\omega_B$, $\omega_R$, $\omega_C$ with a coupling strength quantified by the Rabi frequencies $\Omega_B$,
$\Omega_R$, and $\Omega_C$ respectively. The one-photon detunings of the three lasers are $\Delta_B=\omega_B-\omega_{PS}$,
$\Delta_R=\omega_R-\omega_{PD}$, and
$\Delta_C=\omega_C-\omega_{QS}$. In this work, we focus on the simplest version of this atomic system by neglecting any Zeeman effect, which would otherwise lift the degeneracy of each fine-structure level.

The $N$-type level scheme, formed by the four states and the three laser couplings depicted in Fig.~\ref{fig_N}, is strongly unbalanced. It can be viewed as a $\Lambda$ configuration that is strongly driven by two electric dipole transitions, while being weakly coupled to the $|Q\rangle$ state through the electric quadrupole transition in the limite $\Delta_C\gg \Omega_C$. Despite the weakness of this coupling, the internal-state dynamics of the system can be profoundly modified when a resonance condition involving all three transitions is satisfied.

\begin{figure}[htbp]
\begin{center}
\includegraphics[scale=0.5]{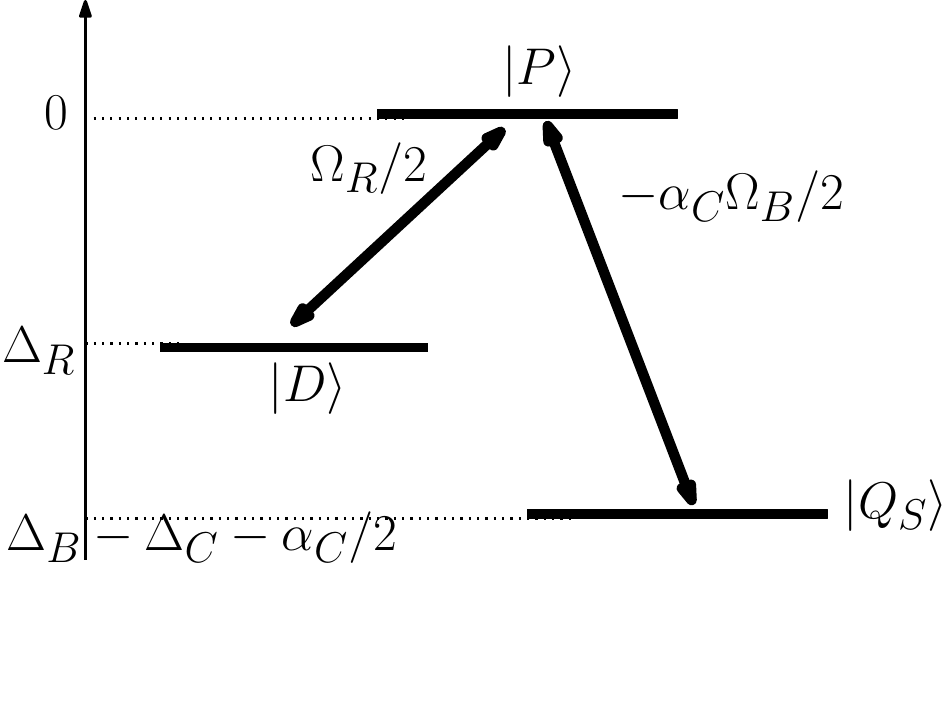} 
\caption{The equivalent $\Lambda$ configuration for 3-photon coherent population trapping.}
\label{fig_lambda}
\end{center}
\end{figure}

This phenomenon is best understood in the dressed-state picture, where the combined 
\{atom+lasers\}
 system is represented in terms of its quantized internal states: each state is associated with $n_B$  photons on transition $B$, $n_R$ photons on transition $R$ and $n_C$ photons on transition $C$ (see Figure \ref{fig_N}-{\bf a} for the level configuration).\\
The atom laser Hamiltonian governing the coherent dynamics of the four-level 
$N$-type system is composed of two parts: the free atomic Hamiltonian and the interaction Hamiltonian describing the couplings induced by the laser fields,
\begin{eqnarray}
H_{AL}=H_0+H_I
\label{hamiltonien initial}
\end{eqnarray}
where 
\begin{equation}
 H_0
=\hbar\Delta_B|S\rangle\langle S|+\hbar(\Delta_B-\Delta_C)|Q\rangle\langle
Q| +\hbar\Delta_R|D\rangle\langle D|
\end{equation}
gives the internal energies of the atomic states in the reference frames of the lasers, while $H_I$ includes the three laser couplings characterized by their Rabi frequencies  $ \Omega_B,  \Omega_R,  \Omega_C$   which are proportional to the local laser electric field amplitudes, whatever the nature of the interaction (electric dipole or quadrupole).
Under the electric dipole and quadrupole interaction approximations, the interaction Hamiltonian can be written as:
\begin{equation}
 \label{Ham}
H_I = \frac{\hbar\Omega_B}{2}|P\rangle\langle S|+\frac{\hbar\Omega_R}{2}|P\rangle\langle D| +\frac{\hbar\Omega_C}{2}|Q\rangle\langle S|+{\rm H.c.}
\end{equation}
Like further explained in \cite{Champenois06},  the dressed laser-coupled subsystem \{$|S\rangle$, $|Q\rangle$\} can be diagonalised at the lowest order of the perturbation in $\alpha_C=\Omega_C/2\Delta_C\ll 1$ and the new eigenstates $|S_Q\rangle$ and $|Q_S\rangle$ are  coherent superposition of the two uncoupled states
\begin{eqnarray}
|S_Q\rangle&=&\textit{N}\left( |S\rangle+\alpha_C |Q\rangle\right)\\ 
|Q_S\rangle&=&\textit{N}\left( |Q\rangle-\alpha_C |S\rangle\right)\label{etat_Qs} 
\end{eqnarray}
(where  \textit{N} is the normalization factor) with eigenfrequencies light-shifted by $\pm \delta_C $ where $\delta_C=\alpha_C \Omega_C/2$.  The new eigenstate $|Q_S\rangle$  is coupled to $|P\rangle$ by the Rabi frequency  $-\alpha_C \Omega_B/2$ (see figure \ref{fig_lambda}). Reference \cite{Champenois06} shows that when $|Q_S\rangle$ and $|D\rangle$ are quasi-resonant ($|\Delta_R|\sim  |\Delta_B-\Delta_C|$) and $\Delta_C\neq 0$, the $|S_Q\rangle$ state is not involved in the dynamics of the internal state.  Including $|D\rangle$, the subsystem $\{|Q_S\rangle, |P\rangle, |D\rangle \}$,  forms a $\Lambda$ configuration where the two feet are stable or metastable as show in figure \ref{fig_lambda}. 
This scheme represents a paradigmatic configuration that gives rise to coherent population trapping (CPT) in a dark state when the two (meta)stable states become degenerate in the dressed-state picture \cite{arimondo96}. Since the effective $\Lambda$ configuration has been obtained by first diagonalizing one of the couplings, the resonance condition leading to CPT now involves three photons instead of two. This three-photon resonance condition can be expressed as
\begin{equation}
\Delta_R=\Delta_B-\Delta_C-\delta_C
\label{accord_D}.
\end{equation}
$\delta_C$ denotes the light shift induced by the quadrupole coupling on the 729nm transition \cite{Champenois06}.
The experimental requirements for achieving a stationary dark resonance impose a stable phase relationship between the three dressing lasers, as is well known in the case of two-photon CPT \cite{vanier05}.

\begin{figure}[hbt]
  \centering
  \includegraphics[scale=0.7]{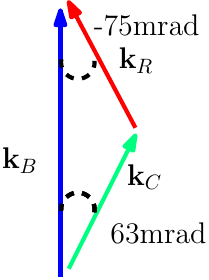}
  \caption{Orientation of the three excitation lasers to obtain $\Delta\textbf{k}$ = 0}\label{arientation lasers}
  \end{figure}

To satisfy this condition, the experiment described in \cite{collombon19b} employs a commercial optical frequency comb (OFC) to transfer phase stability between the three lasers \cite{collombon19}. In this setup, an ultrastable laser at 729~nm developed to excite the electric quadrupole $C$-transition serves as the local reference for the offset-free OFC. The two other lasers driving the $B$- and $R$-transitions are phase-locked to this comb, ensuring that the phases of all three lasers remain strongly correlated and coherent over time. When the Doppler effect due to atomic motion is taken into account, the three-photon resonance condition [Eq.~(\ref{accord_D})] becomes:
\begin{equation}
\Delta_R=\Delta_B-\Delta_C-\delta_C -\Delta\textbf{k}.\textbf{v}
\label{accord_D doppler}
\end{equation}
where \textbf{v} is the velocity of an atom, $\Delta\textbf{k}=\textbf{k$_R$}-\textbf{k$_B$} +\textbf{k$_C$}$ is the relative wave vector of the system and \textbf{k}$_X$ is the wave vector of laser $X$. The lasers involved in the dark resonance condition enable the cancellation of the first-order Doppler effect through a geometric phase-matching of their three wave vectors \cite{champenois07, grynberg76, hong05, ryabtsev11, barker16}. This phase-matching condition is fulfilled when the three excitation lasers are oriented according to the geometry illustrated in Figure~\ref{arientation lasers},
In the following, we assume that this matching condition is satisfied and we neglect any Doppler effect.

The Hamiltonian of the subsystem $\{|Q_S\rangle, |P\rangle, |D\rangle \}$ in the three photons resonance condition (\ref{accord_D})  can be written as
\begin{equation}
H_S=-\hbar\Delta_R|P\rangle\langle P|-\frac{\hbar\alpha_c\Omega_B}{2}|P\rangle\langle Q_S|+\frac{\hbar\Omega_R}{2}|P\rangle\langle D| +{\rm H.c.}
\end{equation} 
This Hamiltonian has the same form as that describing the Stimulated Raman Adiabatic Passage (STIRAP) process in a three-leve $\Lambda$ configuration where the effective Rabi frequencies $-\alpha_c\Omega_B$ and  $\Omega_R$ play the roles of the pump and Stokes fields, respectively \cite{li16}.

In this configuration, the trapping (dark) state corresponds to a coherent superposition of the three stable or metastable dressed states, and can be written as \cite{Champenois06}. The trapping state is a coherent superposition of the three stable and metastable dressed states, and is given by \cite{Champenois06}  
\begin{eqnarray}
|\Psi_{D}\rangle=\textit{N'}\left( \frac{\alpha_C\Omega_B}{\Omega_R} |D\rangle +|Q_S\rangle \right) 
\end{eqnarray}
(\textit{N'} is the normalization factor). This state is not coupled by laser excitation and once trapped in this state, the ions do not emit any photon if the three photon resonance contition is fulfilled. Atomic population can be transfered adiabatically between the two qubit states $|D\rangle$ et $|Q\rangle$ by STIRAP method (see reference \cite{kamsap13} for more details). The general idea behind the population transfer process is to select a time-dependent variation of the system parameters that enables a transition between two metastable states under adiabatic conditions. These conditions require that, for a sufficiently slow temporal evolution of the Hamiltonian, the quantum system remains in its instantaneous eigenstate throughout the process. To achieve the transfer from the initial state $|D\rangle$ to final state $|Q\rangle$, we assume a weak quadrupole coupling such that $\alpha_c$ must be less than 1. Under this assumption, the dressed state $|Q_S\rangle$ can be approximated as $|Q\rangle$, as shown by the relation (\ref{etat_Qs}). Consequently, the time evolution of $\alpha_c\Omega_B/\Omega_R$ must satisfy the following condition:
\begin{equation}
\lim_{t \to 0}  \frac{\alpha_C\Omega_B}{\Omega_R} \gg 1 \hspace{2cm}  \lim_{t \to T_f}  \frac{\alpha_C\Omega_B}{\Omega_R} \ll 1
\end{equation}
where $t=0$ and $T_f$ are initial and final time of transfer. In order to satisfy this condition, we assume the lasers pulses as an adiabatic reference, whith Rabi frequencies evolving according to
\begin{eqnarray}
\Omega_B &=& \frac{ \Omega_0 }{ \alpha_C} 
\exp\left[ -\left( \frac{t-T_f/2+\tau}{\sigma}\right) ^2\right]
 \label{rabi bleu}\\
\Omega_R &=& \Omega_0 \exp\left[ -\left( \frac{t-T_f/2-\tau}{\sigma}\right)^2 \right] \label{rabi rouge}
\end{eqnarray}
where $2\sigma\sqrt{\text{ln2}}$ is the full width at half maximum, $2\tau$ the separation time between the two pulse maxima, and $\Omega_0$ the scaling amplitude. The effective manipulation of quantum information requires very short operation times. However, adiabatic processes are very time-consuming, which is a limiting factor. To address this, we design new modified laser pulse profiles from Gaussian profiles (\ref{rabi bleu}) and (\ref{rabi rouge}).

\section{STIRSAP protocol  in a four-level N-type system} 
To design the STIRSAP protocol, we made some assumptions that are later justified by numerical simulations that take into account the full system. The system is governed by the time-dependent Schr\"odinger equation
\begin{equation}
i\hbar\frac{\partial |\psi (t)\rangle}{\partial t}=H_S  |\psi (t)\rangle
\end{equation}
where $ |\psi (t)\rangle=c_{Q_S}(t) |Q_S\rangle+ c_P(t)  |P\rangle  +c_D(t)|D\rangle$ is the wave functions of the states. We assume à large detuning $\Delta_R\gg(\Omega_R,\alpha_C\Omega_B)$ and we adiabatically eliminate the population of the excited state $|P\rangle$ considering $(\dot{c}_P=0)$ to obtain the effective two levels systems  $\{|Q_S\rangle, |D\rangle \}$ with the following Hamiltonian:
\begin{equation}
\begin{aligned}
H_{eff}=&-\frac{\hbar\Delta_{eff}}{2}|D\rangle\langle D|+\frac{\hbar\Delta_{eff}}{2}|Q_S\rangle\langle Q_S| \\
&+\left( \frac{\hbar \Omega_{eff}}{2}|Q_S\rangle\langle D| +{\rm H.c.}\right) 
\end{aligned}
\label{hamiltonien effectif}
\end{equation}

where 
\begin{eqnarray}
\Delta_{eff}&=&-\frac{\left( \alpha_c \Omega_B\right)^2-\Omega_R^2} {4\Delta_{R}}\label{delta eff}\\
\Omega_{eff}&=&-\frac{\alpha_c \Omega_B\Omega_R}{2\Delta_R}\label{omega eff}
\end{eqnarray}
In order to achieve fast population transfer between $|D\rangle$ and $|Q_S\rangle$  within a short time, we use the counterdiabatic driving given by \cite{drimpak03, berry09, chen10, campo13, li16}:
\begin{equation}
H_{cd}=i\hbar\Sigma_n |\frac{\partial \lambda_n}{\partial t}\rangle\langle \lambda_n |\label{counterdiabatic driven}
\end{equation}
where the eigenstate $ |\lambda_ n\rangle$ of the Hamiltonian (\ref{hamiltonien effectif})   is expressed as $|\lambda_ +\rangle=\sin \theta/2 |D\rangle + \cos\theta/2|Q_S\rangle$ and $|\lambda_ -\rangle=\cos \theta/2 |D\rangle - \sin\theta/2|Q_S\rangle$ where $\tan \theta=-\Omega_{eff}/\Delta_{eff}$. We can calculate the counterdiabatic driving (\ref{counterdiabatic driven}) as \cite{chen10}:
\begin{equation}
H_{cd}=-\frac{i\hbar}{2}\Omega_a|D\rangle\langle Q_S|+ \frac{i\hbar}{2}\Omega_a|Q_S\rangle\langle D|
\label{counterdiabatic}
\end{equation} 
where $\Omega_a=\left( \Omega_{eff}\dot{\Delta}_{eff}-\dot{\Omega}_{eff} \Delta_{eff}\right) /\left( \Delta_{eff}^2+\Omega_{eff}^2\right) $. $\Omega_a$  plays the role of the Rabi frequency for a fast-driving field. Assisted by the counterdiabatic term (\ref{counterdiabatic}), the
system can be driven along the adiabatic path of reference
Hamiltonian (\ref{hamiltonien effectif}) within a short time \cite{li16}. The total Hamiltonian $H=H_{eff}+H_{cd}$, can be expressed as
\begin{equation}
\begin{aligned}
H=&-\frac{\hbar\Delta_{eff}}{2}|D\rangle\langle D|+\frac{\hbar\Delta_{eff}}{2}|Q_S\rangle\langle Q_S|\\
&+\left( \frac{\hbar }{2} \sqrt{\left( \Omega_{eff}^2+\Omega_a^2 \right) } e^{i\varphi}|Q_S\rangle\langle D| +{\rm H.c.}\right)
\end{aligned}
 \label{hamiltonien total effectif}
\end{equation}
where $\varphi=\text{artan}\left( \Omega_a/\Omega_{eff}\right) $. We now apply the unitary transformation: 
\begin{equation}
U(t)=e^{-i\varphi/2} |D\rangle\langle D|+ e^{i\varphi/2} |Q_S\rangle\langle Q_S|
\end{equation}
The unitary transformation means the rotation along $z$ axis, which results in the cancellation of $\sigma_y$ term in the Hamiltonian (\ref{hamiltonien total effectif}) \cite{li16}. By unitary transformation of the Schr\"odinger equation, one can show that the dynamics of the system are described by the new hamiltonian $\tilde{H}_{eff}=U^+HU-i\hbar U^+\dot{U}$,
\begin{equation}
\begin{aligned}
\tilde{H}_{eff}= &-\frac{\hbar\tilde{\Delta}_{eff}}{2}|D\rangle\langle D|+\frac{\hbar\tilde{\Delta}_{eff}}{2}|Q_S\rangle\langle Q_S|\\
&+\left( \frac{\hbar \tilde{\Omega}_{eff}}{2}|Q_S\rangle\langle D| +{\rm H.c.}\right)
\end{aligned}
\label{Hamil tilde}
\end{equation}
where $\tilde{\Delta}_{eff}=\Delta_{eff}+\dot{\varphi}$ and $\tilde{\Omega}_{eff}=\sqrt{\Omega_{eff}^2 + \Omega_{a}^2}$.
Now let us go back to the three level $\Lambda$ system $\{|Q_S\rangle, |P\rangle, |D\rangle \}$ and design the modified pump and Stokes fields by comparing the Hamiltonian (\ref{Hamil tilde}) and (\ref{hamiltonien effectif}). Like Equations (\ref{delta eff}) and (\ref{omega eff}),
we impose
\begin{eqnarray}
\tilde{\Delta}_{eff}&=&-\frac{\left(\alpha_c \tilde{ \Omega}_B\right) ^2-\tilde{\Omega}_R^2}{4\tilde{\Delta}_{R}}\label{delta eff til}\\
\tilde{\Omega}_{eff}&=&-\frac{\alpha_c\tilde{ \Omega}_B\tilde{\Omega}_R}{2\tilde{\Delta}_R}\label{omega eff til}
\end{eqnarray}
where $\tilde{\Omega}_{R}$ and $\tilde{\Omega}_{B}$  are the  modified lasers fields. In order to guarantee large detuning, we should have $\tilde{\Delta}_{R}>>\tilde{ \Omega}_{B,R}$ and we assume $\tilde{\Delta}_{R}=\Delta_{R}$. Then we calculate the modified lasers pulses required to obey equations (\ref{delta eff til}) and (\ref{omega eff til})
\begin{eqnarray}
 \tilde{ \Omega}_B&=&\frac{1}{\alpha_c}\sqrt{-2\Delta_R\left( \sqrt{\tilde{\Delta}_{eff}^2+\tilde{\Omega}_{eff}^2}+\tilde{\Delta}_{eff}\right) }\label{rabi bleu modif}\\
\tilde{ \Omega}_R&=&\sqrt{-2\Delta_R\left( \sqrt{\tilde{\Delta}_{eff}^2+\tilde{\Omega}_{eff}^2}-\tilde{\Delta}_{eff}\right) }\label{rabi rouge modif}.
\end{eqnarray}
We finally obtain newly designed laser fields to drive the state following the dynamics of the systems.  To compare the modified fields (\ref{rabi bleu modif}) and (\ref{rabi rouge modif}) used for STIRSAP to the original fields (\ref{rabi bleu}) and (\ref{rabi rouge}) used for STIRAP in Hamiltonian (\ref{hamiltonien initial}), they are plotted on Fig \ref{compare stirap et stirsap}. The new fields keep the same order of magnitude as the original ones. Note that, under the condition $\alpha_C \ll 1$, the Rabi frequency peak for $\Omega_B$ is much higher than for $\Omega_R$. However, this requirement can become experimentally constraining if it necessitates the use of excessively high laser intensities. It is therefore essential to conduct the analysis using realistic and experimentally accessible Rabi frequencies.

To this end, we now perform numerical simulations to investigate the dynamics of population transfer within the complete four-level system. These simulations allow us to quantify the efficiency of the process, evaluate the role of pulse parameters such as peak Rabi frequencies, and detuning, and to identify optimal conditions for achieving robust population transfer between the two metastable states.

\section{Numerical simulations}

To follow the internal dynamics of this four-level system coupled by three lasers, we solve the master equation for the density matrix $\rho$
\begin{equation} \label{eq_master}
\frac{\partial}{\partial t}\rho=-\frac{\rm i}{\hbar}[H,\rho]+{\cal L}\rho \end{equation}
The term ${\cal L}\rho$ represents the Lindblad super-operator, which accounts for non-Hermitian processes such as spontaneous emission and decoherence. In the present system, we take into account the decay channels originating from the short-lived excited state $|P\rangle$, which can spontaneously decay into the two lower states $|S\rangle$ and $|D\rangle$.

\begin{eqnarray}
{\cal L}_P\rho
&=&-\frac{1}{2}\gamma_P \left(\rho|P\rangle\langle
P|+|P\rangle\langle P|\rho\right) \label{eq_Lp}\\
& &+\beta_{PS}\gamma_P|S\rangle\langle P|\rho|P\rangle\langle S|
+\beta_{PD}\gamma_P|D\rangle\langle P|\rho|P\rangle\langle D|  \nonumber
 \end{eqnarray}
with a branching ratio $\beta_{PS}/\beta_{PD}= 14.6$ for Ca$^{+}$ \cite{safronova11,ramm13} and $\beta_{PS}+\beta_{PD}=1$. The $P_{1/2}$ state lifetime $\gamma_P^{-1}$ used for the simulations is the most recent measured value $6.90$~ns \cite{hettrich15} which is in agreement with very precise calculations \cite{safronova11} that recommend to use $\gamma_P^{-1}=6.87 \pm 0.13$~ns. In a theoretical analysis of this atomic system, the 1-seconde lifetime of the two metastable $D$-states can be ignored \cite{Champenois06}. Nevertheless, with the aim of identifying the optimal experimental parameters, we note that taking these lifetimes into account can have a significant impact. In this work, the spontaneous emission from the metastable states is therefore included through ${\cal L}_D\rho$ :
\begin{eqnarray}
{\cal L}_D\rho
&=&-\frac{1}{2}\gamma_{D_{3/2}} \left(\rho|D\rangle\langle D|+|D\rangle\langle D|\rho\right)+\gamma_{D_{3/2}}|S\rangle\langle D|\rho|D\rangle\langle S| \label{eq_Ld}\\
& & -\frac{1}{2}\gamma_{D_{5/2}} \left(\rho|Q\rangle\langle Q|+|Q\rangle\langle Q|\rho\right)+\gamma_{D_{5/2}}|S\rangle\langle Q|\rho|Q\rangle\langle S|\nonumber
 \end{eqnarray}
 where $\gamma_{D_{3/2}}^{-1}=1195$ms \cite{shao16} and $\gamma_{D_{5/2}}^{-1}=1165$ms \cite{shao17} are the lifetime of the metastable state $|D_{3/2}\rangle$ and $|D_{5/2}\rangle$ respectively.

\begin{figure}[htb]
\centering
\includegraphics[scale=0.22]{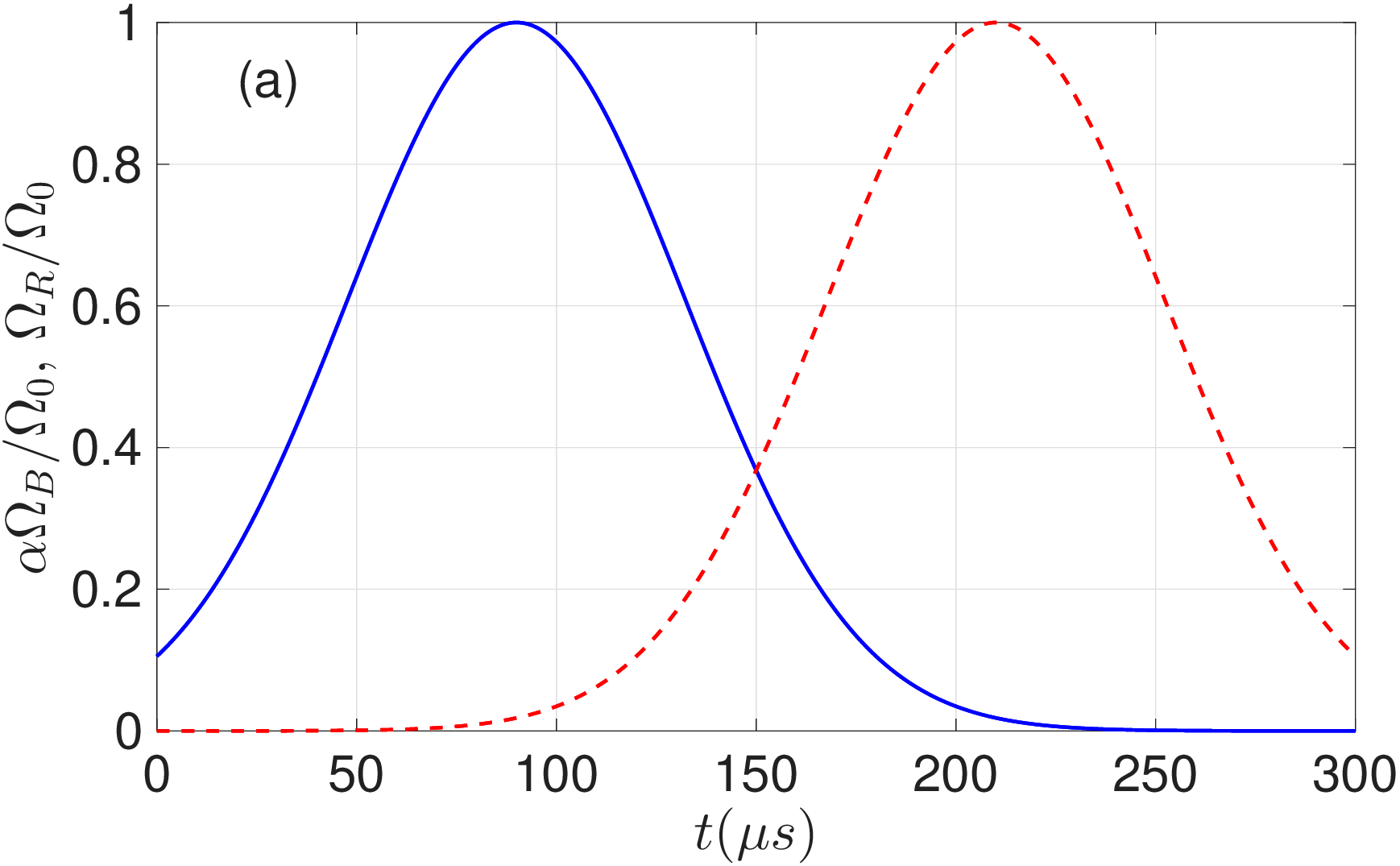}
\includegraphics[scale=0.22]{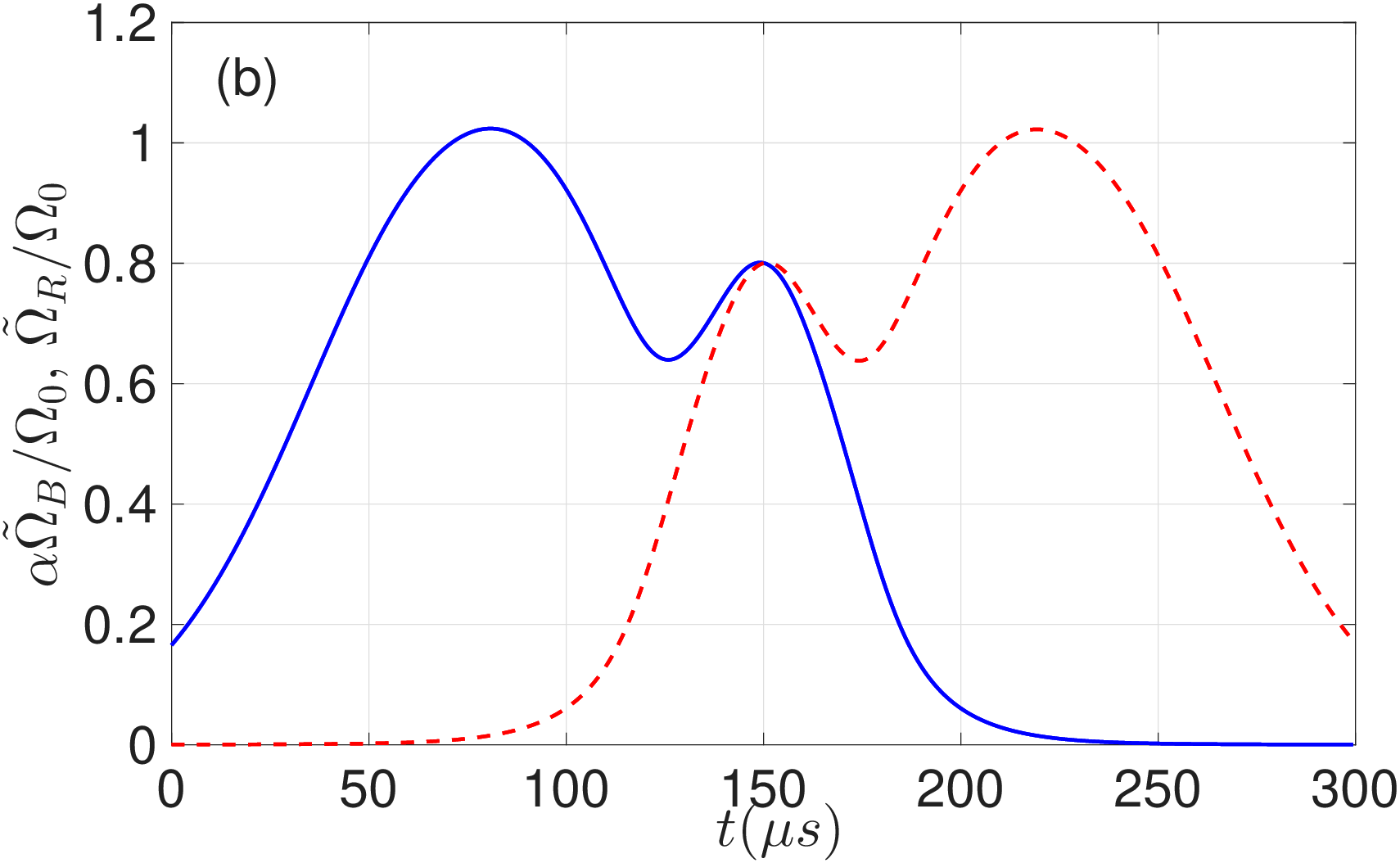}
\includegraphics[scale=0.22]{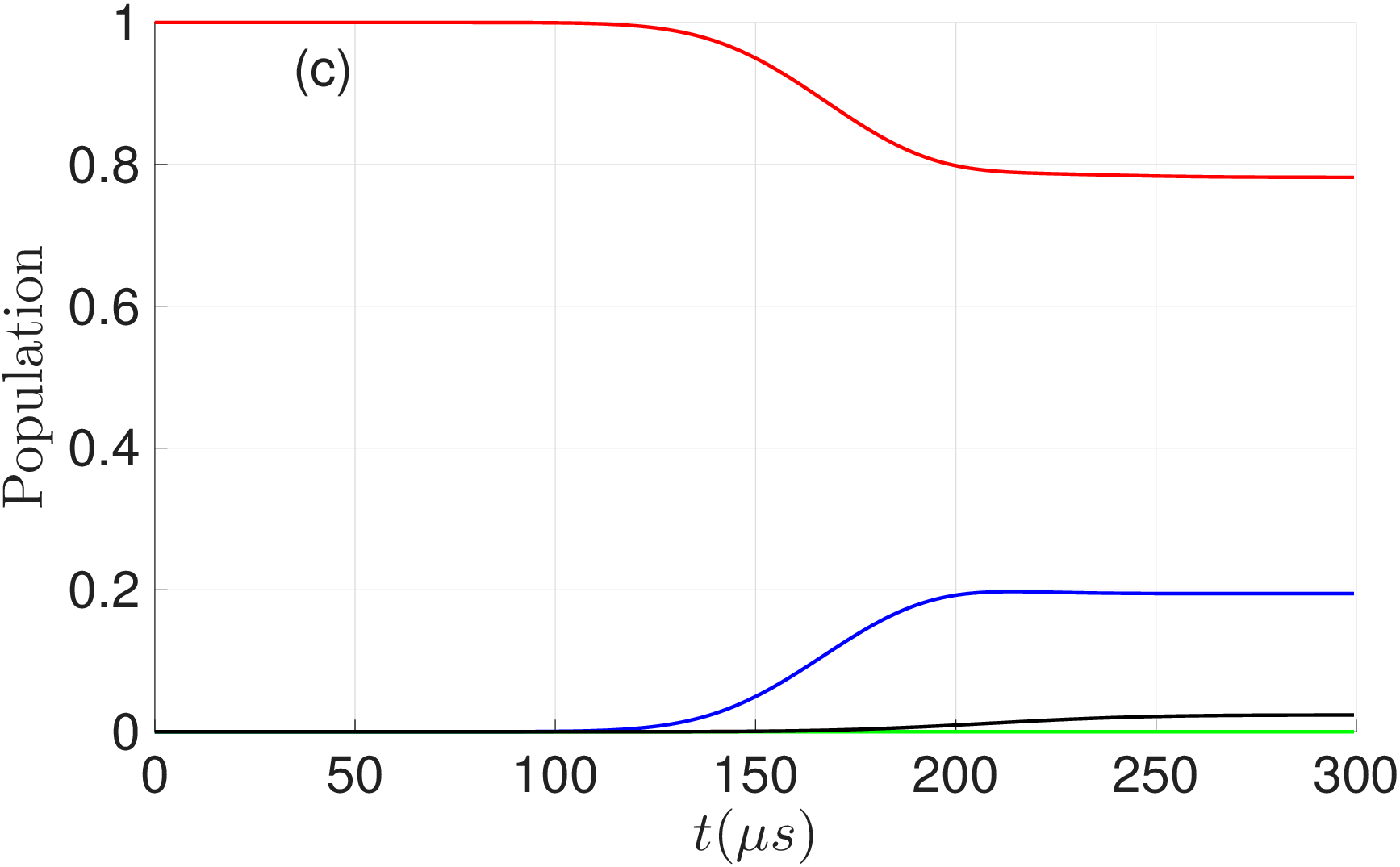}
\includegraphics[scale=0.22]{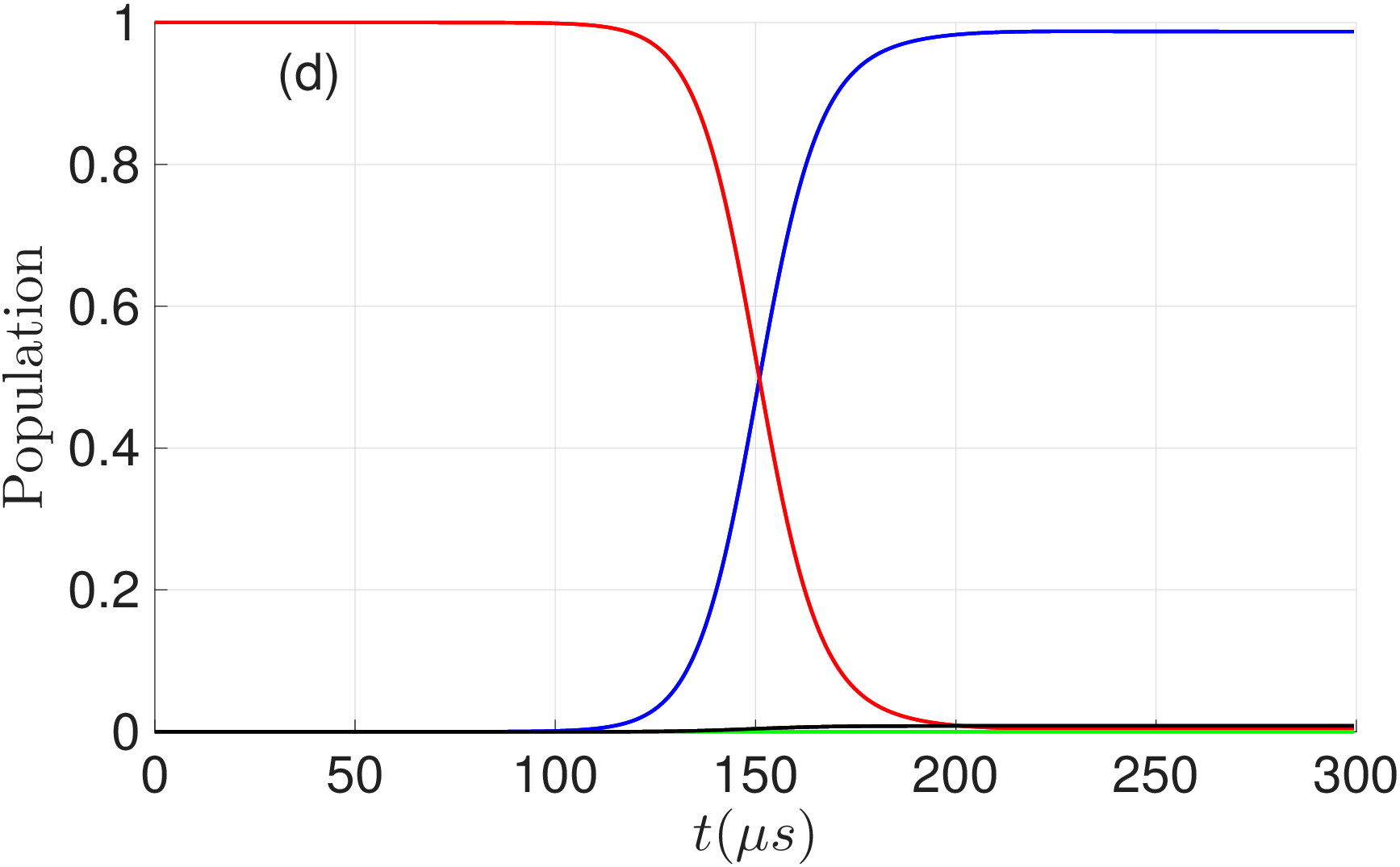}
\caption{Rabi frequencies (\ref{rabi bleu}) and (\ref{rabi rouge}) for STIRAP (a) and (\ref{rabi bleu modif}) and (\ref{rabi rouge modif}) for STIRSAP (b) where the solid blue is $\alpha_C\Omega_B$ en (a) (respectively $\alpha_C\tilde{\Omega}_B$ in (b)) and  the dashed red line is  $\Omega_R$ in (a) (respectively $\tilde{\Omega}_R$ in (b) ). Population dynamics  in $|S\rangle$ state (black curve),  $|Q\rangle$ state (blue curve),  $|D\rangle$ state (red curve) and  $|P\rangle$ state (green curve) for STIRAP (c) and STIRSAP (d), when the initial population are in the $|D\rangle$ state.  Parameters: $\Omega_0/2\pi=$10~MHz, $\Omega_c/2\pi=$5~MHz, $\Delta_C/2\pi=$50~MHz, $\Delta_R/2\pi=-3$~GHz, $T_f=300\mu$s, and $\tau=\sigma=T_f/5$.}
\label{compare stirap et stirsap}
\end{figure}

\begin{figure}[hbt]
 \centering
 \includegraphics[scale=0.24]{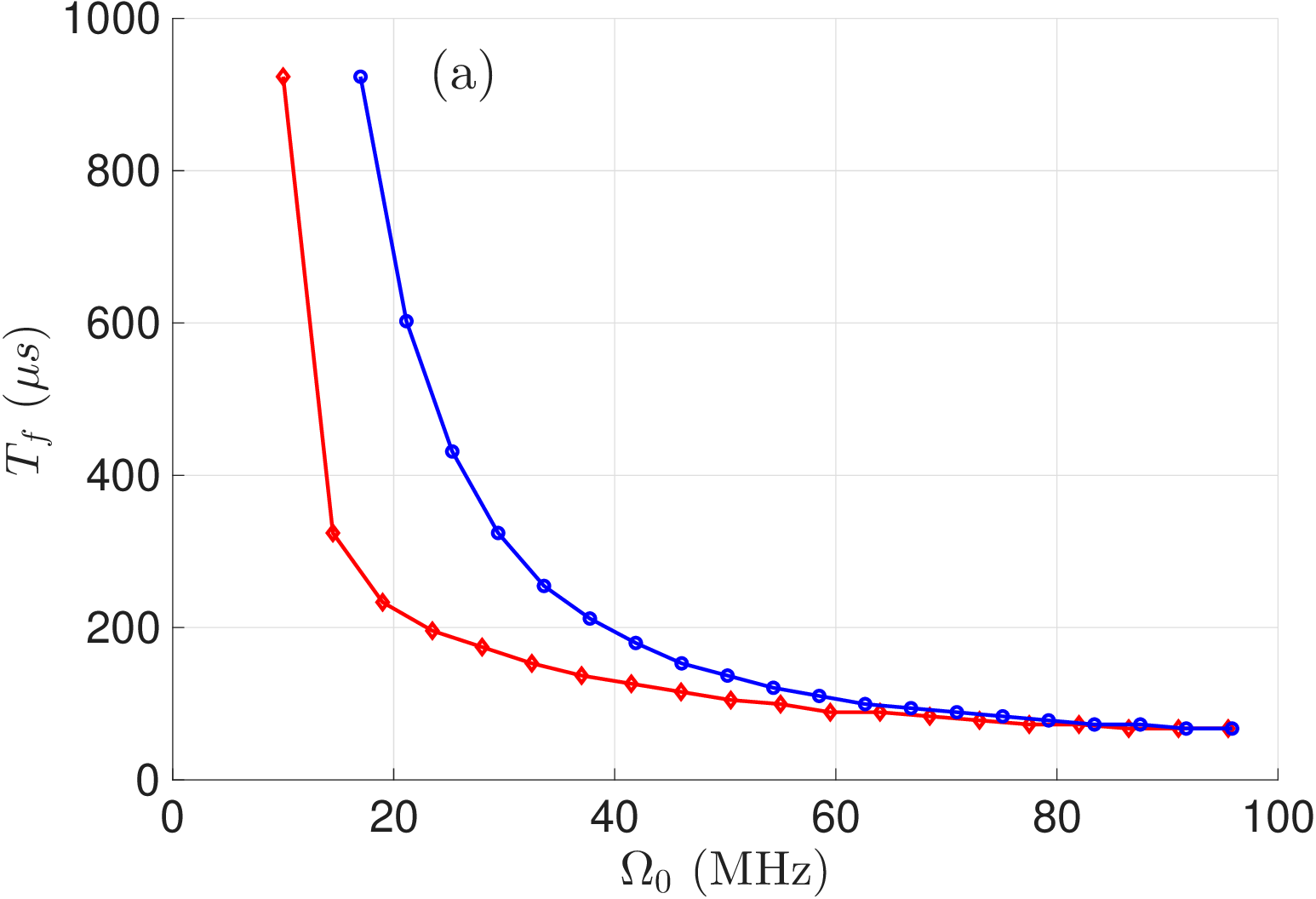}
 \includegraphics[scale=0.24]{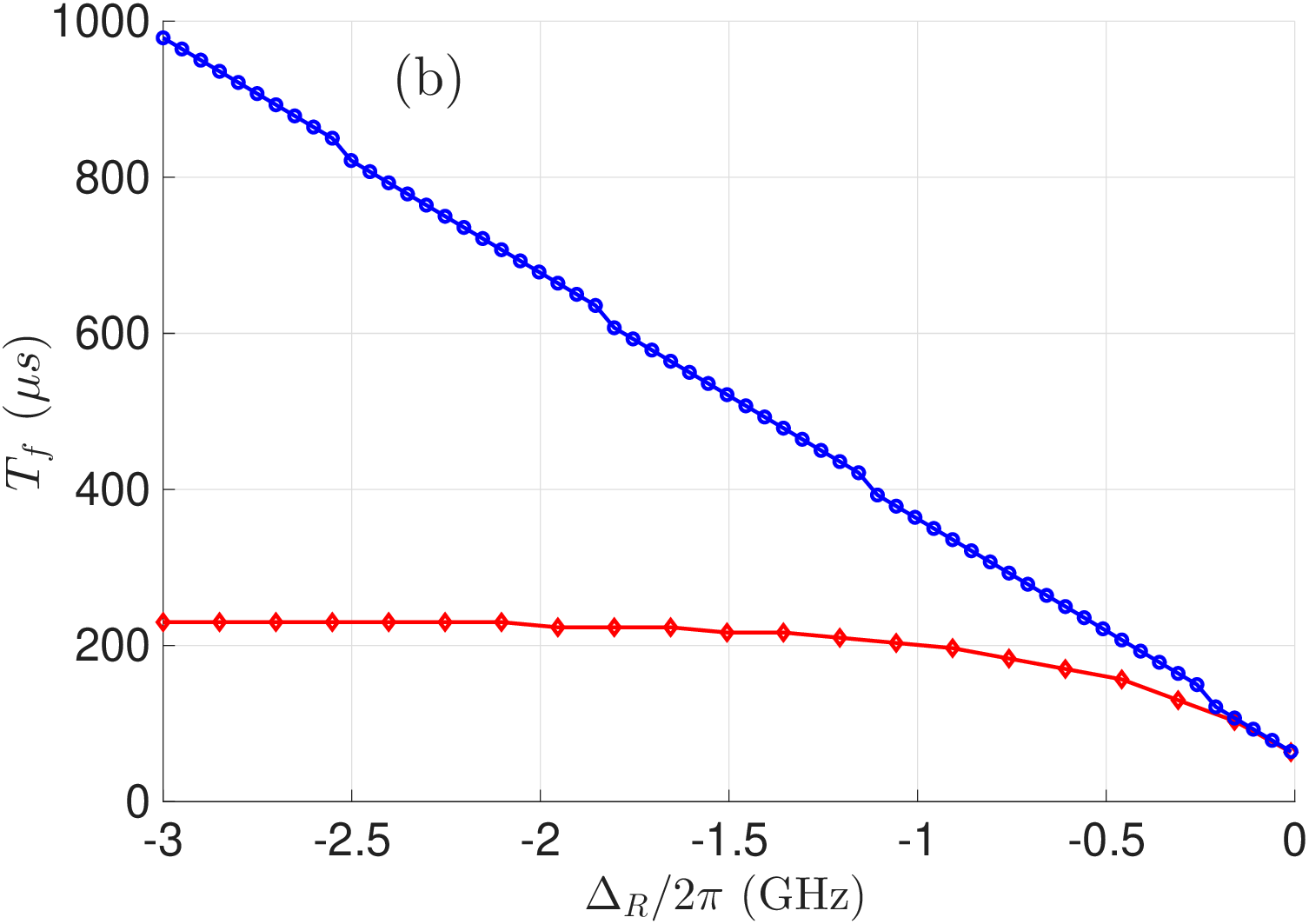}
 \caption{Operation time $T_f$ required to reach a fidelity of 99\% as a function of the peak Rabi frequency $\Omega_0$ (a) and the detuning $\Delta_R$ (b) for STIRAP (blue line) and STIRSAP (red line). Parameters: $\Omega_c/2\pi = 5~\mathrm{MHz}$, $\Delta_C/2\pi = -50~\mathrm{MHz}$, $\tau = \sigma = T_f/5$, and $\Delta_R/2\pi = -2~\mathrm{GHz}$ in (a); $\Omega_0/2\pi = 20~\mathrm{MHz}$ in (b).}
 \label{temps transport}
 \end{figure}

For this work, we assume that the detunings $\Delta_R$ is of the order of GHz to guarantee large detuning ($\Delta_R\gg\Omega_0$) and choose $\alpha_C=\Omega_C/2\Delta_C\ll 1$.  Note that, $\Omega_0$ and $\Omega_0/\alpha_C$ are the amplitudes of Rabi frequencies $\Omega_R$ and $\Omega_B$ respectively.  We compare the dynamics using the original field given by  equation (\ref{rabi bleu}) and (\ref{rabi rouge}) as shown in the figure \ref{compare stirap et stirsap}-a for STIRAP (see figure \ref{compare stirap et stirsap}-c), and  the modified field given by (\ref{rabi bleu modif}) and (\ref{rabi rouge modif}) as shown in figure  \ref{compare stirap et stirsap}-b for STIRSAP (see figue \ref{compare stirap et stirsap}-d), for $\alpha_C=0.05$.
We observe that the shapes of the modified fields differ from the original Gaussian profiles by adding an extra local maximum to the time evolution. Nevertheless, the pulses remain smooth and experimentally feasible for implementation with cold atoms \cite{du16}. Within short interaction times, the population transfer from state $|D\rangle$ to $|Q\rangle$ is incomplete in the case of the standard STIRAP protocol, whereas complete population transfer is achieved when using the STIRSAP technique. These results are consistent with those previously reported for three-level systems in a $\Lambda$ configuration and are particularly relevant in the context of quantum information processing, where fast and high-fidelity state transfer is essential. As states $|S\rangle$ and $|P\rangle$ remain unpopulated throughout the process, the pulses can be stopped at any time to produce any combination of $|D\rangle$ and $|Q\rangle$ states.  To quantify the difference in operation time for transfers using STIRAP and STIRSAP, we evaluated the time required to achieve 99\% fidelity as a function of amplitude $\Omega_0$ for a given set of parameters. The results are shown in Figure \ref{temps transport}-a.  Overall, we note that the operation time $T_f$ required to reach 99\% is shorter for STIRSAP than for STIRAP for given parameters. The gain is noticeable for small values of $\Omega_0$.  For example, with $\Omega_0/2\pi$=20~MHz, the operation times are $T_f=220 \mu$s and $700 \mu$s for STIRSAP and STIRAP respectively, i.e. the STIRSAP time is more than three times shorter. This difference becomes greater as $\Omega_0$ decreases. Conversely, for very large $\Omega_0$ both techniques converge to similar operation times, consistent with the adiabatic limit where transfer speed is no longer limited by diabatic transitions. Since adiabatic transfer typically requires large Rabi frequencies, which are experimentally constrained by available laser power, STIRSAP offers a practical route to achieve fast and efficient population transfer while minimising the required optical intensity.
Furthermore, Fig.~\ref{temps transport}-b compares the operation times obtained with STIRAP (blue line) and STIRSAP (red line) for different detuning $\Delta_R$. For STIRAP, the operation time increases approximately linearly with the detuning $|\Delta_R|$, whereas it remains nearly constant for STIRSAP when $|\Delta_R|$ exceeds $1~\mathrm{GHz}$. This behavior highlights the robustness of STIRAP against large detuning and provides quantitative support for our initial assumption.

We further numerically quantified the photon emission from the $|P\rangle$  state during the transfer process with relation $N_\mathrm{photon}=\gamma_P\int^\infty_0 \langle P|\rho|P\rangle dt$. The number of emitted photons decreases as the detuning $\Delta_R$ increases. In particular, for $|\Delta_R/2\pi| > 1~\mathrm{GHz}$, the average number of emitted photons remains below $0.04$, whereas it reaches approximately $0.28$ at $|\Delta_R/2\pi| = 10~\mathrm{MHz}$. This behavior also indicates that the small-detuning regime is more affected by spontaneous-emission losses, which we now explicitly acknowledge as a limitation of the approach. Therefore, the reduction in photon emission indicates that larger detunings effectively suppress spontaneous-emission losses by minimizing the population of the intermediate state and make the dynamics effectively coherent. These results justify the wave-function description assumed in deriving the counterdiabatic Hamiltonian in the case of large detuning\\

\begin{figure}[hbtp]
 \centering
 \includegraphics[scale=0.25]{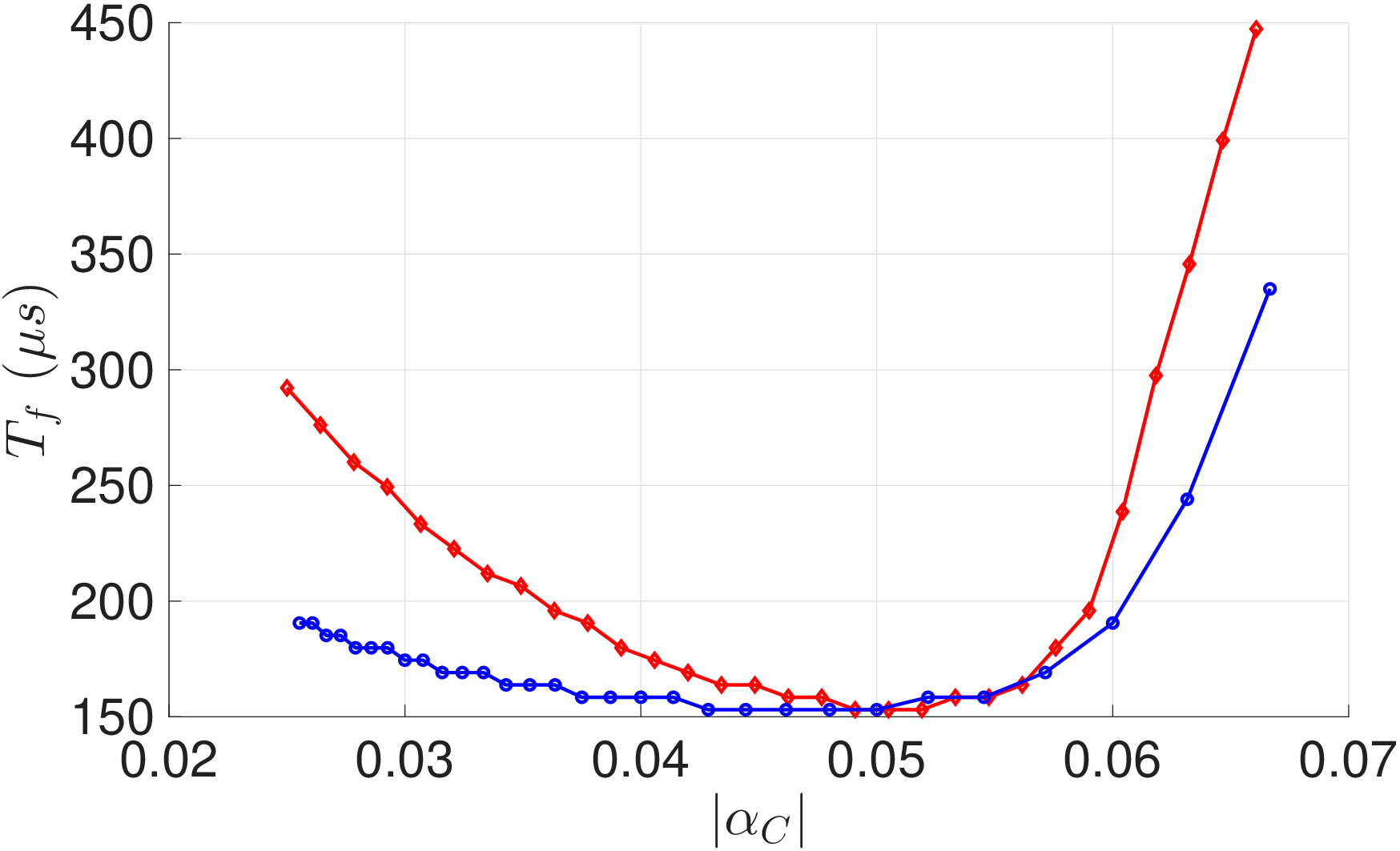}
 \caption{Operation time $T_f$ to reach 99\% fidelity versus $\alpha_C=\Omega_C/2\Delta_C$ for $\Delta_C/2\pi=-100$MHz (blue curve) when we make $\Omega_C$ vary and $\Omega_C/2\pi=10$MHz (red curve) when  $\Delta$ vary. $\Omega_0/2\pi=$20MHz, $\Delta_R/2\pi=$-2GHz, and $\tau=\sigma=T_f/5$ \label{duree_operation_ac}}
 \end{figure}

 Unlike the three-level system, our system relies on an additional coupling with parameter $\alpha_C=\Omega_C/2\Delta_C$, as demonstrated above. This parameter is likely to alter the dynamics of the system.  To optimise the speed and efficiency of the transfer, we analyse its influence. Figure \ref{duree_operation_ac} shows the operating time required for 99\% transfers as a function of $\alpha_C$.  The red curve is plotted with a constant value of -100~MHz for $\Delta_C/2\pi$, and the blue curve with a constant value of 10~MHz for $\Omega_C/2\pi$.

In both cases, we observe  that, there is an optimal value of $\alpha_C = 0.05\pm 0.005$ which minimises transfer time. Beyond this value of $\alpha_C$, transfer times increase, indicating inefficiency. Indeed, it has also been demonstrated that transfer is most efficient when $\alpha_C = 0.05$ in the case of STIRAP \cite{kamsap13}.  When $\alpha_C$  is much less than 0.05, the coupling becomes negligible and the system behaves like a three-level system. Conversely, when $\alpha_C$ is much greater than 0.05, the approximation $\alpha_C \ll 1$ is invalid and the numerical results do not match the expected behaviour. Figure \ref{duree_operation_ac} shows that the optimal value for $\alpha_C$  allows a tolerance of 10\%. This tolerance and the choice of large one-photon detunings  reduce their sensitivity to the Doppler effect. For example, for fixed $\Omega_C/2\pi=10$~MHz,  $\alpha_C = 0.05\pm 0.005$ results in $\Delta_C/2\pi=-100 \pm 10$~MHz. This tolerance for the one-photon detuning corresponds to a Doppler shift induced by a velocity of  7.29~ms$^{-1}$ which is 20 times larger than the mean squared velocity reached by Doppler laser cooling on Ca$^+$ ion. The STIRAP and STIRSAP methods are thus very robust methods relative to the Doppler effect. With the optimal value of $\alpha_C$, the peak Rabi frequency for $\Omega_B$ is 20 times greater than for $\Omega_R$.  While this value is relatively high, it remains achievable if a reasonable choice is made for $\Omega_R$.

 \section{Conclusion}
 
In this study, we have demonstrated the applicability of the STIRSAP protocol to a four-level atomic system in an $N$-configuration, enabling accelerated population transfer between two metastable states. A distinctive feature of this configuration is the presence of a three-photon resonance condition that gives rise to coherent population trapping. This condition effectively eliminates the first-order Doppler effect, thereby rendering the system insensitive to thermal motion. The system was first reduced to three effective levels in a $\Lambda$ configuration, analogous to previously studied cases, and subsequently an effective two-level system through adiabatic elimination. This reduction enabled the application of counterdiabatic driving, followed by a unitary transformation that modifies the pump and Stokes fields without introducing additional couplings. As a result, the adiabatic trajectory is followed exactly, while diabatic transitions are completely suppressed. \\
 Numerical simulations were performed to analyze the influence of system parameters and to compare transfer times between STIRAP and STIRSAP schemes. The results show that, for identical parameters STIRSAP achieves significantly shorter population transfer times or lower laser intensities  than conventional STIRAP for the same transfer duration. Moreover, the quadrupole coupling strength between $|S\rangle$ and $|Q\rangle$ characterized by the parameter $\alpha_C$ possesses an optimal value that maximizes transfer efficiency while minimizing operation time. The robustness of the protocol allow to tolerate Doppler shift on one photon-detuning compatible with Doppler laser cooling. In summary, the STIRSAP technique emerges as a promising and experimentally feasible approach for rapid, high-fidelity quantum state transfer in multi-level atomic systems without the requirement for sub-Doppler cooling techniques. This work contributes to advancing coherent control methods and quantum information processing, offering a robust framework for implementing fast quantum operations beyond the adiabatic limit.

 \section*{acknowledgments}
The authors thank Prof. Dr. Caterina Cocchi from Friedrich-Schiller-Universit\"at Jena, for her collaboration and useful exchanges. MRK would like to thank CNRS for financial support, through the DSCA Africa project, during the completion of this work.

\section*{Declarations}

\begin{itemize}
\item Funding: This work was supported by the "Centre National de la Recherche Scientifique" (CNRS) France through the DSCA Africa project.
\item Conflict of interest: The author states that there is no conflict of interest.
\item Competing interests: The authors declare no competing interests
\item Data availability: No datasets were generated or analysed during the current study.
\item Author contribution: All of the authors contributed to the design of the research project. MRK developed the theoretical framework and simulations, and wrote the initial draft of the article. SEN, MDM and MEA verified and validated the theoretical calculations and simulations. They also read and edited the article. CKS, TBE and ABM approved the project and proofread the manuscript. CC designed the project, validated the scientific approach, reviewed and supervised the work.
\end{itemize} 
 
 \bibliographystyle{apsrev4-2}
\bibliography{bibliotrap}
%\bibliography{your_bibfile} % Upload .bib file separately to arXiv

\end{document}